\documentstyle[prd,aps,floats,tighten,epsfig,graphicx]{revtex}
\newcommand{\beq}{\begin{equation}}
\newcommand{\eeq}{\end{equation}}
\newcommand{\bea}{\begin{eqnarray}}
\newcommand{\ena}{\end{eqnarray}}
\newcommand{\etal}{{\it et al.}}
\newcommand{\ie}{{\it i.e.}}
\newcommand{\lsim}{\mathrel{\mathop{\kern 0pt \rlap
{\raise.2ex\hbox{$<$}}}
\lower.9ex\hbox{\kern-.190em $\sim$}}}
\newcommand{\gsim}{\mathrel{\mathop{\kern 0pt \rlap
{\raise.2ex\hbox{$>$}}}
\lower.9ex\hbox{\kern-.190em $\sim$}}}

\newcommand{\puis}[1]{$^{#1}$}

\newcommand{\app}[3]{Astropart.\ Phys.\ {\bf #1}, #3 (#2)}

\newcommand{\hepph}[1]{{\tt hep-ph/#1}}
\newcommand{\astroph}[1]{{\tt astro-ph/#1}}
\newcommand{\prep}[3]{Phys.\ Rep.\ {\bf #1}, #3 (#2)}
\newcommand{\plb}[3]{Phys.\ Lett.\ B\ {\bf #1}, #3 (#2)}
\newcommand{\npb}[3]{Nucl.\ Phys.\ B\ {\bf #1}, #3 (#2)}

\renewcommand{\apj}[3]{Astrophys.\ J.\ {\bf #1}, #3 (#2)}
\newcommand{\aeta}[3]{Astron.\ {\&}\ Astrophys.\ {\bf #1}, #3 (#2)}
\renewcommand{\prl}[3]{Phys.\ Rev.\ Lett. {\bf #1}, #3 (#2)}
\renewcommand{\prd}[3]{Phys.\ Rev.\ D\ {\bf #1}, #3 (#2)}

\newcommand{\rnc}[3]{Riv.\ Nuovo\ Cim.\ {\bf #1}, #3 (#2)}
\newcommand{\zfpc}[3]{Z.\ Phys.\ C\ {\bf #1}, #3 (#2)}
\newcommand{\href}[2]{#1}

\newcommand{\DFLUX}
{\mbox{$\rm m^{-2} \; s^{-1} \; sr^{-1} \; GeV^{-1}$}} 

\newcommand{\pbar}{\mbox{$\bar{\rm p}$}}
\newcommand{\nbar}{\mbox{$\bar{\rm n}$}}
\newcommand{\dbar}{\mbox{$\bar{\rm D}$}}
\newcommand{\kX}{\mbox{$\vec{k_{\rm X}}$}}

\newcommand{\kpbar}{\mbox{$\vec{k_{\pbar}}$}}
\newcommand{\knbar}{\mbox{$\vec{k_{\nbar}}$}}
\newcommand{\kdbar}{\mbox{$\vec{k_{\dbar}}$}}
\newcommand{\kdif}{\mbox{$\vec{\Delta}$}}
\newcommand{\STOT}{\mbox{$\sigma_{\rm tot}$}}
\newcommand{\STOTPP}{\mbox{$\sigma_{\rm p - p}^{\rm tot}$}}
\newcommand{\mpt}{\mbox{$m_{\rm p}$}}

\newcommand{\md}{\mbox{$m_{\dbar}$}}
\newcommand{\pcoal}{\mbox{$P_{\rm coal}$}}

\newcommand{\Ep}{\mbox{$E_{\rm p}$}}
\newcommand{\Epbar}{\mbox{$E_{\pbar}$}}
\newcommand{\Edbar}{\mbox{$E_{\dbar}$}}

\newcommand{\mC}{\mbox{$m_{\chi}$}}

\begin{document}

\title{Antideuterons as a Signature of Supersymmetric Dark Matter
\setcounter{footnote}{3}
\footnote{Report Nos. IFIC/99-9, FTUV/99-9, LAPTH--722/99}}
\vskip 1.cm
\author{
Fiorenza Donato$^{\rm a}$
\setcounter{footnote}{0}
\footnote{E--mail: donato@lapp.in2p3.fr, fornengo@flamenco.ific.uv.es,
salati@lapp.in2p3.fr},
Nicolao Fornengo$^{\rm b}$ and Pierre Salati$^{\rm a,c}$
}
\vskip 0.5cm
\address{
\begin{flushleft}
a) Laboratoire de Physique Th\'eorique LAPTH, BP110, F-74941
Annecy-le-Vieux Cedex, France.\\
b) Instituto de F\'isica corpuscular - C.S.I.C.,
Departamento de F\'isica Te\`orica,
Universitat de Val\'encia, C./ Dr Moliner 50,\\
E--46100 Burjassot, Val\'encia, Spain.\\
c) Universit\'e de Savoie, BP1104 73011 Chamb\'ery Cedex,
France.
\end{flushleft}
}
\maketitle
\vskip 0.5cm
\centerline{Draft Version of \today}

\vskip 0.5cm
\begin{abstract}
Once the energy spectrum of the secondary component
is well understood, measurements of the antiproton
cosmic--ray flux at the Earth will be a powerful way to indirectly
probe for the existence of supersymmetric relics in the galactic halo. 
Unfortunately, it is still spoilt by considerable theoretical
uncertainties. As shown in this work, searches for
low--energy antideuterons appear in the mean time as a
plausible alternative, worth being explored.
Above a few GeV/n, a dozen spallation antideuterons should
be collected by the future AMS experiment on board ISSA.
For energies less than $\sim$ 3 GeV/n, the {\dbar} spallation
component becomes negligible and may be supplanted by a
potential supersymmetric signal. If a few low--energy antideuterons
are discovered, this should be seriously taken as a clue for the
existence of massive neutralinos in the Milky Way.
\end{abstract}
\pacs{98.70.Sa, 95.35.+d, 14.20.-c, 14.80.Ly}
\vskip 1.cm

\section{Introduction.}
\label{sec:introduction}

Cosmic ray fluxes are about to be measured with unprecedented
precision both by balloon borne detectors and by space instruments.
The various ongoing experiments are also hunting for traces
of antimatter in the cosmic radiation. The BESS collaboration
\cite{bess_1}
plans to push the limit on the $\bar{\rm He}/{\rm He}$ ratio
down to $10^{-8}$ whereas the AMS spectrometer should reach a
sensitivity of $\sim 10^{-9}$ once it is installed on the International
Space Station Alpha (ISSA) \cite{ams}.
The search for antinuclei has profound cosmological implications.
The discovery of a single antihelium or anticarbon would actually
be a smoking gun for the existence of antimatter islands in our
neighborhood. However, light antinuclei, mostly antiprotons but also
antideuterons, are actually produced in our Galaxy as secondaries.
They result from the interaction of high--energy cosmic--ray protons
with the interstellar gas of the Milky Way disk. In a previous analysis,
Chardonnet {\etal} \cite{chardonnet97} have estimated the flux of
antideuterium {\dbar} and antihelium $\bar{\rm ^{3}He}$ secondaries.
The {\dbar} signal is very weak but may marginally be detected by AMS
on board ISSA. The case of antihelium is, at least for the moment, hopeless.

The dark matter of the Milky Way could be made mostly of elementary
particles such as the heavy and neutral species predicted by supersymmetry.
The mutual annihilations of these relics, potentially concealed in the
halo of our Galaxy, would therefore produce an excess in the cosmic radiation
of gamma rays, antiprotons and positrons. In particular, supersymmetric
antiprotons should be abundant at low energy, a region where the flux of
{\pbar} secondaries is a priori negligible. There is quite an excitement trying
to extract from the observations a possible {\pbar} exotic component
which would signal the presence of supersymmetric dark matter
in the Galaxy. Unfortunately, it has been recently realized
\cite{bottino98,bergstrom99,bieber99} that a few processes add up
together to flatten out, at low energy, the spectrum of secondary
antiprotons.
Ionisation losses as well as inelastic but non-annihilating scatterings
on the hydrogen atoms of the galactic disk result into the decrease of the
antiproton energy. The low--energy tail of the {\pbar} spectrum is
replenished by the more abundant population from higher energies.
That effect is further strengthened by solar modulation which also shifts
the energy spectrum towards lower energies.
As a result of these effects, the secondary {\pbar}'s are much more
abundant at low energy than previously thought. Disentangling
an exotic supersymmetric contribution from the conventional
component of spallation antiprotons may turn out to be a very
difficult task. The antiproton signal of supersymmetric dark matter
is therefore in jeopardy.

Antideuterons, \ie, the nuclei of antideuterium, are free from such problems.
As explained in Sect.~\ref{sec:coalescence}, they form when an antiproton
and an antineutron merge together. The two antinucleons must be
at rest with respect to each other in order for fusion to take place
successfully. For kinematic reasons, a spallation reaction creates very
few low--energy particles. Low-energy secondary antideuterons are even
further suppressed. Energy loss mechanisms are also less efficient in shifting
the antideuteron energy spectrum towards low energies. The corresponding
interstellar (IS) flux is derived in Sect.~\ref{sec:secondary}, for energies in
the range extending from 0.1 up to 100 GeV/n. It reaches a maximum of
$2-5 \times 10^{-8}$ {\dbar} {\DFLUX}
for a kinetic energy of $\sim$ 4 GeV/n. A dozen of
secondary antideuterons should be collected by the
AMS/ISSA experiment.

On the other hand, supersymmetric {\dbar}'s are manufactured at rest
with respect to the Galaxy. In neutralino annihilations, antinucleons are
predominantly produced with low energies. This feature is further
enhanced by their subsequent fusion into antideuterons, hence
a fairly flat spectrum for supersymmetric antideuterium nuclei
as shown in Sect.~\ref{sec:susy}. 
Below a few GeV/n, secondary antideuterons are quite
suppressed with respect to their supersymmetric partners. That
low--energy suppression is orders of magnitude more effective
for antideuterons than for antiprotons. This makes
cosmic--ray antideuterons a much better probe of supersymmetric
dark matter than antiprotons.

Unfortunately, antideuteron fluxes are quite small with respect
to {\pbar}'s. We nevertheless show in Sect.~\ref{sec:conclusion}
that a significant portion of the supersymmetric parameter space may
be explored by measuring the cosmic--ray {\dbar} flux at low energy.
In particular, an AMS/ISSA caliber experiment should reach a
sensitivity of $4.8 \times 10^{-8}$ {\dbar} {\DFLUX}
at solar minimum, pushing it down to
$3.2 \times 10^{-8}$ {\dbar} {\DFLUX}
at solar maximum, for a modulated energy of 0.24 GeV/n.

\section{Production of antideuterons.}
\label{sec:coalescence}

At this point, our goal is to derive the cross section for the production
of antideuterons. The processes at stake are both the spallation of a cosmic--ray 
high--energy proton on an hydrogen atom at rest and the annihilation
of a neutralino pair. The number $d{\cal N}_{\rm X}$ of particles ${\rm X}$
-- antinucleons or antideuterons -- produced in a single reaction and whose
momenta are $\kX$, is related to the differential production cross section
through
\beq
d{\cal N}_{\rm X} \; = \; {\displaystyle \frac{1}{\STOT}} \,
{d^{3} \sigma_{\rm X}}(\sqrt{s} , \kX)
\;\; ,
\eeq
where $\STOT$ denotes the total cross section for the process under 
scrutiny -- spallation reaction or neutralino annihilation. The total
available energy is $\sqrt{s}$. The corresponding differential probability
for the production of ${\rm X}$ is defined as
\beq
d{\cal N}_{\rm X} \; = \; {\cal F}_{\rm X} (\sqrt{s} , \kX) \, d^{3} \kX
\;\; .
\eeq
For each of the processes under concern, the differential probability
for the production of an antiproton or an antineutron may be derived.
The calculation of the probability for the formation of an antideuteron
can now proceed in two steps. We first need to estimate the probability for
the creation of an antiproton--antineutron pair. Then, those antinucleons
merge together to yield an antinucleus of deuterium.

As explained in Ref.~\cite{chardonnet97}, the production of two antinucleons is
assumed to be proportionnal to the square of the production of one of them.
The hypothesis that factorization of the probabilities holds is fairly well
established at high energies. For spallation reactions, however, the bulk
of the antiproton production takes place for an energy $\sqrt{s} \sim 10$ GeV
which turns out to be of the same order of magnitude as the antideuteron mass.
Pure factorization should break in that case as a result of energy conservation.
It needs to be slightly adjusted. We have therefore assumed that the center
of mass energy available for the production of the second antinucleon is reduced
by twice the energy carried away by the first antinucleon
\beq
{\cal F}_{\pbar , \nbar} (\sqrt{s} , \kpbar , \knbar) \; = \;
{\displaystyle \frac{1}{2}} \,
{\cal F}_{\pbar} (\sqrt{s} , \kpbar) \,
{\cal F}_{\nbar} (\sqrt{s} - 2 E_{\pbar} , \knbar) \; + \;
\left( \kpbar \leftrightarrow \knbar \right)
\;\; .
\eeq
Once the antiproton and the antineutron are formed, they combine
together to give an antideuteron with probability
\beq
{\cal F}_{\dbar} (\sqrt{s} , \kdbar) \, d^{3} \kdbar \; = \;
{\displaystyle \int} \, d^{3} \kpbar \, d^{3} \knbar \;
{\cal C}(\kpbar , \knbar) \;
{\cal F}_{\pbar , \nbar} \left( \sqrt{s} , \kpbar , \knbar \right) \;\; .
\label{coalescence_1}
\eeq
The summation is performed on those antinucleon configurations for which
\beq
\kpbar + \knbar \; = \; \kdbar \;\; .
\eeq
The coalescence function ${\cal C}(\kpbar , \knbar)$ describes the probability
for a $\pbar - \nbar$ pair to yield by fusion an antideuteron. That function
depends actually on the difference
$\kpbar - \knbar = 2 \kdif$ between the antinucleon momenta so that
relation (\ref{coalescence_1}) may be expressed as
\beq
{\cal F}_{\dbar} (\sqrt{s} , \kdbar) \; = \;
{\displaystyle \int} \, d^{3} \kdif \; {\cal C}(\kdif) \;
{\cal F}_{\pbar , \nbar} \left( \sqrt{s} ,
\kpbar \, = \, \frac{\kdbar}{2} + \kdif ,
\knbar \, = \, \frac{\kdbar}{2} - \kdif \right) \;\; .
\label{coalescence_2}
\eeq
An energy of $\sim 3.7$ GeV is required to form by spallation an
antideuteron whereas the binding energy of the latter is
$B \sim 2.2$ MeV. The coalescence function is therefore strongly
peaked around
$\kdif = \vec{0}$ and expression (\ref{coalescence_2}) simplifies into
\beq
{\cal F}_{\dbar} (\sqrt{s} , \kdbar) \; \simeq \; \left\{
{\displaystyle \int} \, d^{3} \kdif \; {\cal C}(\kdif) \right\} \;
{\cal F}_{\pbar , \nbar} \left( \sqrt{s} ,
\kpbar \, = \, \frac{\kdbar}{2} , \knbar \, = \, \frac{\kdbar}{2}
\right) \;\; ,
\eeq
where the probability for the formation of the $\pbar - \nbar$ pair
has been factored out. The term in brackets may be estimated in the
rest frame of the antideuteron through the Lorentz invariant term
\beq
{\displaystyle \int} \, \frac{E_{\dbar}}{E_{\pbar} \, E_{\nbar}} \,
d^{3} \kdif \, {\cal C}(\kdif) \; \simeq \;
\left( \frac{\md}{m_{\pbar} \, m_{\nbar}} \right) \,
\left(\frac{4}{3} \pi \pcoal^{3} \right) \;\; .
\eeq 
In that frame, the antinucleons merge together if the momentum of the
corresponding two--body reduced system is less than some critical value
$\pcoal$. That coalescence momentum is the only free parameter of our
factorization and coalescence scheme. As shown in Ref.~\cite{chardonnet97},
the resulting antideuteron production cross section in proton--proton
collisions is well fitted by this simple one--parameter model. A value of
$\pcoal = 58$ MeV has been derived, not too far from what may be naively 
expected from the antideuteron binding energy, \ie,
$\sqrt{\mpt \, B} \sim 46$ MeV.

The differential probability with which an antiproton is produced during
a proton--proton collision is related to the corresponding Lorentz invariant
cross section through
\beq
{\STOTPP} \, E_{\pbar} \, {\cal F}_{\pbar} (\sqrt{s} , \kpbar) \; = \;
\left. E_{\pbar} \, \frac{d^{3} \sigma}{d^{3} \kpbar} \right|_{\rm LI}
\;\; .
\eeq
The latter is experimentally well known. It is fairly well fitted by the Tan and Ng's
parametrization \cite{TanNg82} which has been used here. Assuming that
the invariance of isospin holds, the antineutron production cross section is
equal to its antiproton counterpart.
The Lorentz invariant cross section for the production of antideuterons
resulting from the impact of a high--energy cosmic--ray proton on a
proton at rest has been derived by Chardonnet \etal \cite{chardonnet97}
who showed that
\bea
E_{\dbar} \,
{\displaystyle \frac{d^3 \sigma_{\dbar}}{d^{3} \kdbar}} & = &
\left( {\displaystyle \frac{\md}{m_{\pbar} \, m_{\nbar}}} \right) \,
\left({\displaystyle \frac{4}{3}} \pi \pcoal^{3} \right) \times
{\displaystyle \frac{1}{2 \STOTPP}} \times \nonumber \\
& \times &
\left\{
E_{\pbar} {\displaystyle \frac{d^{3} \sigma_{\pbar}}{d^{3} \kpbar}}
\left( \sqrt{s} , \kpbar \right) \,
E_{\nbar} {\displaystyle \frac{d^{3} \sigma_{\nbar}}{d^{3} \knbar}}
\left( \sqrt{s} - 2 E_{\pbar} , \knbar \right)
\; + \;  \left( \kpbar \leftrightarrow \knbar \right)
\right\}
\;\; .
\label{LI_dbar_production}
\ena
The corresponding differential cross section obtains from the summation, in
the galactic frame, of the Lorentz invariant production cross section
(\ref{LI_dbar_production})
\beq
{\displaystyle \frac{d \sigma_{\rm p H \to \dbar}}{d \Edbar}}
\left\{ \Ep \to \Edbar \right\} \; = \;
2 \pi \; k_{\dbar} \;
{\displaystyle \int_{0}^{\theta_{\rm max}}} \,
\left. E_{\dbar} \, \frac{d^{3} \sigma}{d^{3} \kdbar} \right|_{\rm LI} \;
d \left( - \cos \theta \right) \;\; .
\label{integral_production_lab}
\eeq
In that frame, $\theta$ denotes the angle between the momenta of the incident
proton and of the produced antideuteron. It is integrated up to a maximal value
$\theta_{\rm max}$ set by the requirement that, in the center of mass
frame of the reaction, the antideuteron energy $E^{*}_{\dbar}$ cannot exceed
the bound
\beq
E^{*}_{\dbar \, {\rm max}} \; = \;
{\displaystyle \frac{s \, - \, 16 {\mpt}^{2} \, + \, {\md}^{2}}{2 \sqrt{s}} }
\;\; .
\eeq
The integral (\ref{integral_production_lab}) is performed at fixed
antideuteron energy ${\Edbar}^{2} = {\md}^{2} + {k_{\dbar}}^{2}$.

In the case of a neutralino annihilation, the differential multiplicity
for antiproton production may be expressed as
\beq
{\displaystyle \frac{dN_{\pbar}}{d \Epbar}} \; = \;
{\displaystyle \sum_{\rm F , h}} \, B_{\rm \chi h}^{\rm (F)} \,
{\displaystyle \frac{dN_{\pbar}^{\rm h}}{d \Epbar}} \;\; .
\eeq
The annihilation proceeds, through the various final states F, towards
the quark or the gluon h with the branching ratio $B_{\rm \chi h}^{\rm (F)}$.
Quarks or gluons may be directly produced when a neutralino pair annihilates.
They may alternatively result from the intermediate production of a Higgs or
gauge boson as well as of a top quark. Each quark or gluon h generates in turn
a jet whose subsequent fragmentation and hadronization yields the antiproton
energy spectrum ${dN_{\pbar}^{\rm h}} / {d \Epbar}$.
Because neutralinos are at rest with respect to each other, the probability
to form, say, an antiproton with momentum $\kpbar$ is essentially  isotropic
\beq
{\displaystyle \frac{dN_{\pbar}}{d \Epbar}}(\chi + \chi \to \pbar + \ldots)
\; = \;
4 \pi \, k_{\pbar} \, \Epbar \, {\cal F}_{\pbar}(\sqrt{s} = 2 \mC , \Epbar)
\;\; .
\eeq
Applying the factorization--coalescence scheme discussed above leads
to the antideuteron differential multiplicity
\beq
{\displaystyle \frac{dN_{\dbar}}{d \Edbar}}\; = \;
\left( {\displaystyle \frac{4 \, \pcoal^{3}}{3 \, k_{\dbar}}} \right)
\,
\left( {\displaystyle \frac{\md}{m_{\pbar} \, m_{\nbar}}} \right)
\;
{\displaystyle \sum_{\rm F , h}} \, B_{\rm \chi h}^{\rm (F)} \,
\left\{
{\displaystyle \frac{dN_{\pbar}^{\rm h}}{d \Epbar}}
\left( \Epbar = \Edbar / 2 \right)
\right\}^{2} \;\; .
\label{dNdbar_on_dEdbar_susy}
\eeq
It may be expressed as a sum, extending over the various quarks and
gluons h as well as over the different annihilation channels F, of the
square of the antiproton differential multiplicity. That sum is weighted
by the relevant branching ratios. The antineutron and antiproton differential
distributions have been assumed to be identical.
The hypothesis that factorization holds is certainly conservative. The 
antinucleons which merge together to create an antideuteron are produced
in the same quark or gluon jet. Their momenta are not isotropically distributed
with respect to each other. They tend to be more aligned than what has been
assumed here, with a larger chance to generate an antideuteron. However,
our analysis is meant to be conservative.

\section{The detection of spallation antideuterons.}
\label{sec:secondary}

\begin{figure*}[htb!]
\centerline{
\resizebox{0.7\textwidth}{!}
{\includegraphics*[1.5cm,6.5cm][18.5cm,23.cm]{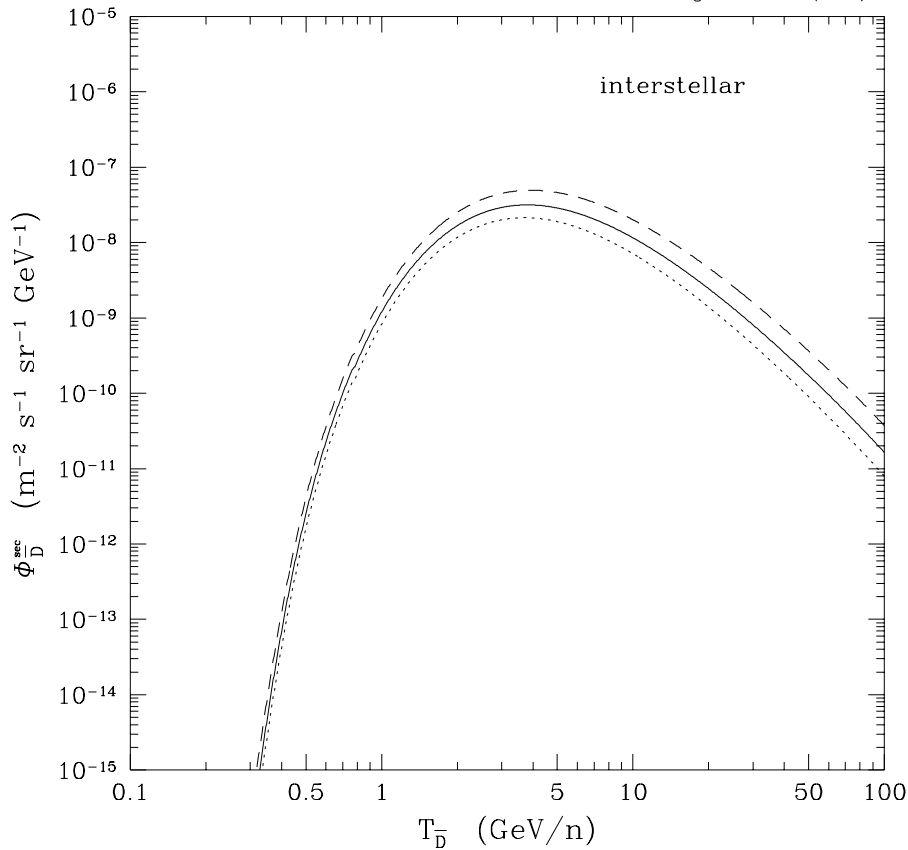}}
}
\caption{
The IS secondary flux of antideuterons, expressed
in units of \DFLUX, is presented as a function of kinetic
energy per nucleon. The solid curve corresponds to the median value
of the cosmic--ray proton spectrum, as derived by Bottino \etal [4].
The dashed and dotted lines respectively stand for
the maximal and minimal values of the primary proton flux
from which the antideuterons originate.
}
\label{fig:dbar_sec_is}
\end{figure*}

\begin{figure*}[htb!]
\centerline{
\resizebox{0.7\textwidth}{!}
{\includegraphics*[1.5cm,6.5cm][18.5cm,23.cm]{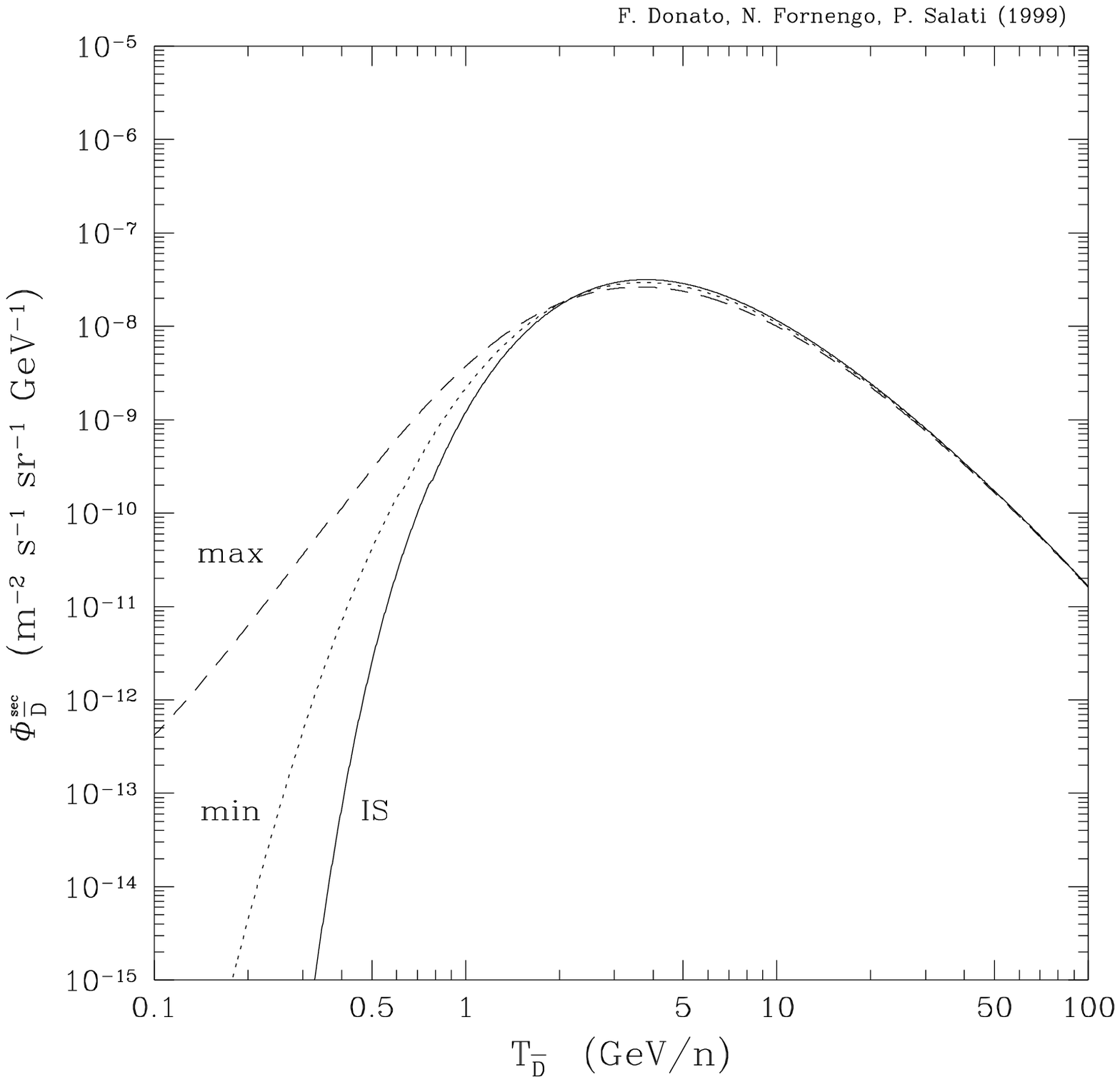}}
}
\caption{
The median IS spectrum of Fig.~\ref{fig:dbar_sec_is} (solid curve)
has been modulated at solar maximum (dashed line) and minimum
(dotted line).
}
\label{fig:dbar_sec_solmod}
\end{figure*}

As suggested by Parker, the propagation of cosmic--rays inside the Galaxy is
strongly affected by their scattering on the irregularities of magnetic fields.
This results into a diffusive transport. In the following, we will assume an
isotropic diffusion with an empirical value for the diffusion coefficient.
Our Galaxy can be reasonably well modelled by a thin disk of atomic and
molecular hydrogen, with radius $R \sim 20$ kpc and thickness $\sim$
200 pc. This gaseous ridge is sandwiched between two diffusion regions
which act as confinement domains as a result of the presence of irregular 
magnetic fields. They extend vertically up to $\sim 3$ kpc apart from the
central disk. That two-zone diffusion model is in good agreement with the
observed primary and secondary nuclei abundances \cite{webber92}.

Assuming a steady regime, the propagation of cosmic--ray antideuterons
within the Milky Way is accounted for by the diffusion equation
\begin{equation}
- \, K \Delta \psi_{\dbar} \, + \,
\Gamma_{\dbar} \, \psi_{\dbar}\; + \;
\frac{\partial}{\partial E} \left\{ b(E) \psi_{\dbar}\right\}
\; = \; q_{\dbar}^{\rm sec} \;\; ,
\label{diffusion_dbar_disk}
\end{equation}
where \mbox{$\psi_{\dbar}$} is the density of antideuterons per unit of
volume and per unit of energy.
In the left--hand side of relation (\ref{diffusion_dbar_disk}), the first term
describes the diffusion of the particles throughout the galactic magnetic
fields. The coefficient $K$ is derived from measurements of the light element
abundances in cosmic--rays. It is constant at low energies, but beyond a
critical value of ${\cal R}_{0} = 1$ GV, it raises with rigidity $\cal{R}$
like
\beq
K({\cal{R}}) \; = \; K_{0} \,
\left( 1 + \frac{\cal{R}}{\cal R}_{0} \right)^{0.6} \;\; ,
\label{DIFFUSION_K}
\eeq
where $K_{0} = 6 \times 10^{27}$ cm$^{2}$ s$^{-1}$. It is assumed to be
essentially independent of the nature of the species that propagate throughout
the Galaxy.
The second term acounts for the destruction of antideuterons through their
interactions, mostly annihilations, with the interstellar medium.
Antideuterons may also undergo fragmentation if they survive annihilation.
In that case, they are broken apart as most of the cosmic--ray nuclei.
The total collision rate is given by
\beq
\Gamma_{\dbar} \; = \;
\sigma_{\dbar \, {\rm H}} \; v_{\dbar} \; n_{\rm H} \;\; ,
\label{pbar_collision}
\eeq
where $\sigma_{\dbar \, {\rm H}}$ is the total antideuteron interaction
cross section with protons \cite{sigma_dbar_H}, $v_{\dbar}$ denotes the
velocity and $n_{\rm H} = 1$ cm\puis{-3} is the average hydrogen density
in the thin matter disk.
The last term in the left--hand side of relation (\ref{diffusion_dbar_disk})
stands for the energy losses undergone by antideuterons as they diffuse in
the galactic ridge. The rate
$b (\Edbar) = \dot{\Edbar}$
at which the antideuteron energy varies is essentially set by the ionization
losses which the particle undergoes as it travels through interstellar gas.
This mechanism yields the following contribution to the energy loss rate
\begin{equation}
b_{\rm \, ion} (\Edbar) \; = \;
- \, 4 \pi \, r_{e}^{2} \; m_{e} c^{2} \; n_{\rm H} \;
{\displaystyle \frac{c}{\beta} } \;
\left\{ \ln \left( \frac{2 \, m_{e} c^{2}}{E_{0}} \right) \, + \,
\ln \left( \beta^{2} \gamma^{2} \right) \, - \, \beta^{2} \right\}.
\end{equation}
In molecular hydrogen, the ionization energy $E_{0}$ has been set equal
to 19.2 eV; here $\gamma = \Edbar / \md$. The classical radius of the
electron is denoted by $r_{e}$ and the electron mass is $m_{e}$.
In the case of antiprotons, it was realized
\cite{TanNg82,bottino98,bergstrom99} that the dominant energy
loss mechanism is actually their inelastic, but non--annihilating,
interactions with interstellar protons. The latter are excited towards
resonant states and hence absorb part of the antiproton energy. In the {\pbar}
frame, an incident proton kicks off the antiproton at rest, transfering
some of its kinetic energy. In the case of antideuterons, however, 
such a process is no longer possible. In the {\dbar} frame, the impinging
proton cannot transfer energy without destroying the antideuteron
whose binding energy -- $B \sim 2.2$ MeV -- is much smaller than the
typical kinetic energies at stake. That is why fragmentation generally
dominates the interactions of cosmic--ray nuclei with interstellar matter.
Accordingly, the resulting destruction occurs at fixed energy per nucleon.
%
In the right--hand side of the diffusion Eq.~(\ref{diffusion_dbar_disk}),
the production rate $q_{\dbar}^{\rm sec}$ of the spallation antideuterons
involves a convolution over the incident cosmic--ray proton energy
spectrum $\psi_{\rm p}$ of the differential production cross
section (\ref{integral_production_lab})
\beq
q_{\dbar}^{\rm disk}(\Edbar) \; = \;
{\displaystyle \int_{\Edbar}^{+ \infty}} \; d\Ep \;
\psi_{\rm p}(\Ep) \, n_{\rm H} \, v_{\rm p} \;
{\displaystyle \frac{d \sigma_{\rm p H \to \dbar}}{d \Edbar}}
\left\{ \Ep \to \Edbar \right\}
\;\; .
\label{sec_source}
\eeq

The differential energy distribution $\psi_{\dbar}$ of secondary
antideuterons is determined by solving Eq.~(\ref{diffusion_dbar_disk}).
We have followed the standard approach which may be found in 
Ref.~\cite{berezinskii90}. At the edge of the domain where the cosmic--rays
are confined, the particles escape freely, the diffusion becomes inefficient 
and densities vanish. This provides the boundary conditions for solving 
Eq.~(\ref{diffusion_dbar_disk}). Then, because the problem is axisymmetric,
the various cosmic--ray distributions may be expanded as series of Bessel
functions of zeroth order. Details may be found in
Refs.~\cite{bottino98,chardonnet96}. The secondary antideuteron
interstellar flux finally obtains from the differential energy spectrum
\beq
\Phi_{\dbar}^{\rm sec} \; = \;
\frac{1}{4 \pi} \, \psi_{\dbar} \, v_{\dbar}
\;\; .
\eeq

The interstellar (IS) flux of spallation antideuterons is presented in
Fig.~\ref{fig:dbar_sec_is} as a function of the kinetic energy per nucleon.
As explained in Bottino \etal \cite{bottino98}, the IS proton flux is 
still uncertain around $\sim$ 20--100 GeV, an energy range that
contributes most to the integral~(\ref{sec_source}). We have borrowed
the parametrization
\beq
\Phi_{\rm p}^{\rm IS} \; = \; A \, \beta \,
\left(
{\displaystyle \frac{\Ep}{1 \, {\rm GeV}}}
\right)^{- \alpha} \;\; .
\label{proton_flux}
\end{equation}
The median IS proton flux corresponds to a normalization factor of
$A = 15,950$ protons {\DFLUX} with a spectral index of $\alpha = 2.76$.
The normalization factor $A$ has been varied from $12,300$ (minimal)
up to $19,600$ protons {\DFLUX} (maximal). Accordingly, the minimal
and maximal IS proton fluxes respectively correspond to the spectral
indices $\alpha = 2.61$ and $2.89$. In Fig.~\ref{fig:dbar_sec_is}, the solid 
curve features the IS secondary antideuterons generated from the 
median proton spectrum. The maximal (dashed line) and minimal
(dotted line) distributions delineate the band within which the spallation
antideuteron signal lies. The flux reaches a maximum value
comprised between 2.1 and $4.9 \times 10^{-8}$ {\dbar} {\DFLUX} for
a kinetic energy of $\sim$ 4 GeV/n.
The antideuteron spectrum sharply drops below a few GeV/n.
Remember that in the galactic frame, the production threshold is
17 {\mpt}. When a high--energy cosmic--ray proton impinges
on an hydrogen atom at rest, the bulk of the resulting antiprotons and
antineutrons keep moving, with kinetic energies $\sim 10 - 20$ GeV.
For kinematical reasons, the production of antinucleons at rest with
respect to the Galaxy is extremely unprobable. The manufacture of a
low--energy antideuteron is even more unprobable. It actually requires
the creation of both an antiproton and an antineutron at rest. The momenta
need to be aligned in order for fusion to succesfully take place. Low--energy
antideuterons produced as secondaries in the collisions of high--energy
cosmic--rays with the interstellar material are therefore extremely scarce,
with a completely depleted energy spectrum below $\sim$ 1 GeV/n.
Energy losses tend to shift the antideuteron spectrum towards lower
energies with the effect of replenishing the low--energy tail with the
more abundant species which, initially, had a higher energy. This
process tends to slightly soften the strong decrease of the low--energy
antideuteron spectrum. The effect is nevertheless mild. Remember that
in the case of antiprotons, it is actually the inelastic but non--annihilating
interactions which considerably flatten the {\pbar} distribution. Ionizations
losses are not enough to significantly affect the energy spectrum.
The IS secondary antideuterons are therefore extremely depleted
below $\sim$ 1 GeV/n. The spallation background is negligible
in the region where supersymmetric {\dbar}'s are expected to be
most abundant. This feature makes the detection of low--energy
antideuterons an interesting signature of the presence of supersymmetric
relics in the Galaxy.

In Fig.~\ref{fig:dbar_sec_solmod}, the median IS {\dbar} spectrum (solid curve) 
has been modulated at solar maximum (dashed line) and solar minimum
(dotted line). We have applied the forced field approximation \cite{perko}
to estimate the effect of the solar wind on the cosmic--ray energies
and fluxes. For the energies at stake, this amounts to simply shift the
IS energy of a nucleus $N$, with charge $Z$ and atomic number $A$, by a
factor of $Z e \Phi$. The solar modulation parameter $\Phi$ has the same
dimensions as a rigidity or an electric potential. The Earth ($\oplus$) and
interstellar (IS) energies, {\em per nucleon}, are therefore related by
\beq
E_{N}^{\oplus} / A \; = \;
E_{N}^{\rm IS} / A \, - \, \left| Z \right| e \Phi / A \;\; .
\eeq
In Perko's approximation, antinuclei are affected in just the same way
as nuclei. Their energy decreases as they penetrate inside the heliomagnetic
field. Once the momenta at the Earth $p_{N}^{\oplus}$ and at the boundaries
of the heliosphere $p_{N}^{\rm IS}$ are determined, the flux modulation
ensues
\beq
{\displaystyle
\frac{\Phi_{N}^{\oplus} \left( E_{N}^{\oplus} \right)}
{\Phi_{N}^{\rm IS}  \left( E_{N}^{\rm IS} \right)}}
\; = \;
\left\{ {\displaystyle \frac{p_{N}^{\oplus}}{p_{N}^{\rm IS}}} \right\}^{2} \,
\;\; .
\eeq
Antideuterons undergo an energy loss, {\em per nucleon}, half that of protons
and antiprotons. At solar minimum (maximum) the modulation parameter
$\Phi$ has been set equal to 320 MV (800 MV) \cite{bottino98}. The energy shift
is larger at solar maximum than at solar minimum. Once modulated, the sharply
decreasing IS antideuteron distribution tends therefore to be flatter at solar
maximum as is clear in Fig.~\ref{fig:dbar_sec_solmod}. We estimate that a total of
12--13 secondary antideuterons may be collected by the AMS collaboration
during the space station stage, in the energy range extending up to 100 GeV/n.
These antideuterons correspond to IS energies in excess of $\sim$ 3 
GeV/n, a region free from the effects of solar modulation. This result
takes into account the geomagnetic suppression as discussed in
Sect.~\ref{sec:conclusion}.

\section{The supersymmetric antideuteron signal.}
\label{sec:susy}

\begin{figure*}[h!]
\centerline{
\resizebox{0.7\textwidth}{!}
{\includegraphics*[1.5cm,6.5cm][18.5cm,23.cm]{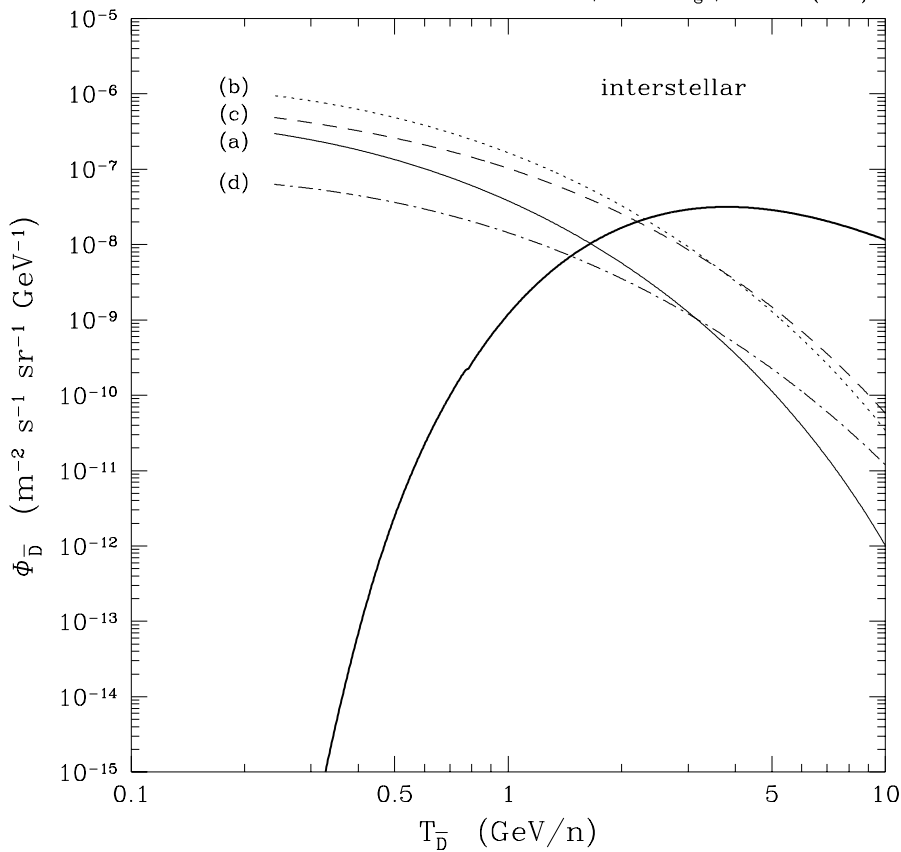}}
}
\caption{
The IS flux of secondary antideuterons (heavier solid curve)
decreases at low energy whereas the energy spectrum of the
antideuterons from supersymmetric origin tends to flatten.
The four cases of table~\ref{table:susy} are respectively featured
by the solid (a), dotted (b), dashed (c) and dot-dashed (d) curves.
}
\label{fig:dbar_prim_is}
\end{figure*}

\begin{table}[h!]
\[
\begin{array}{|c|c|c|c|c|c|c|c|} \hline
{\rm case} &
m_{\chi} &
P_{g} (\%) &
\Omega_\chi h^{2} &
\Phi_{\pbar}^{\rm min} \left( 0.24 \; {\rm GeV} \right) &
\Phi_{\dbar}^{\rm min} \left( 0.24 \; {\rm GeV/n} \right) &
\Phi_{\dbar}^{\rm max} \left( 0.24 \; {\rm GeV/n} \right) &
N_{\dbar}^{\rm max}
\\
& & & & & & &
\\
\hline \hline
a & 36.5 & 96.9 & 0.20  & 1.2 \times 10^{-3} & 1.0 \times 10^{-7} &
2.9 \times 10^{-8} & 0.6\\
b & 61.2 & 95.3 & 0.13 & 3.9 \times 10^{-3} & 3.5 \times 10^{-7}&
1.1 \times 10^{-7} & 2.9 \\
c & 90.4 & 53.7 & 0.03 & 1.1 \times 10^{-3} & 1.8 \times 10^{-7} &
6.1 \times 10^{-8} & 2.0 \\
d & 120 & 98.9 & 0.53  & 2.9 \times 10^{-4} & 2.5 \times 10^{-8}&
8.6 \times 10^{-9} & 0.3  \\
\hline \hline
\end{array} \]
\caption{
These four cases illustrate the richness of the supersymmetric
parameter space. There is no obvious correlation between the
antiproton and antideuteron Earth fluxes with the neutralino mass
$ m_{\chi}$. Case (c) is a gaugino-higgsino mixture and still
yields signals comparable to those of case (a), yet a pure gaugino.
Antideuteron fluxes are estimated at both solar minimum and
maximum, for a modulated energy of 0.24 GeV/n. The last column
features the corresponding number of {\dbar}'s which AMS on board
ISSA can collect below 3 GeV/n.
}
\label{table:susy}
\end{table}

\begin{figure*}[h!]
\centerline{
{\resizebox{0.5\textwidth}{!}
{\includegraphics*[1.5cm,6.5cm][18.5cm,23.cm]{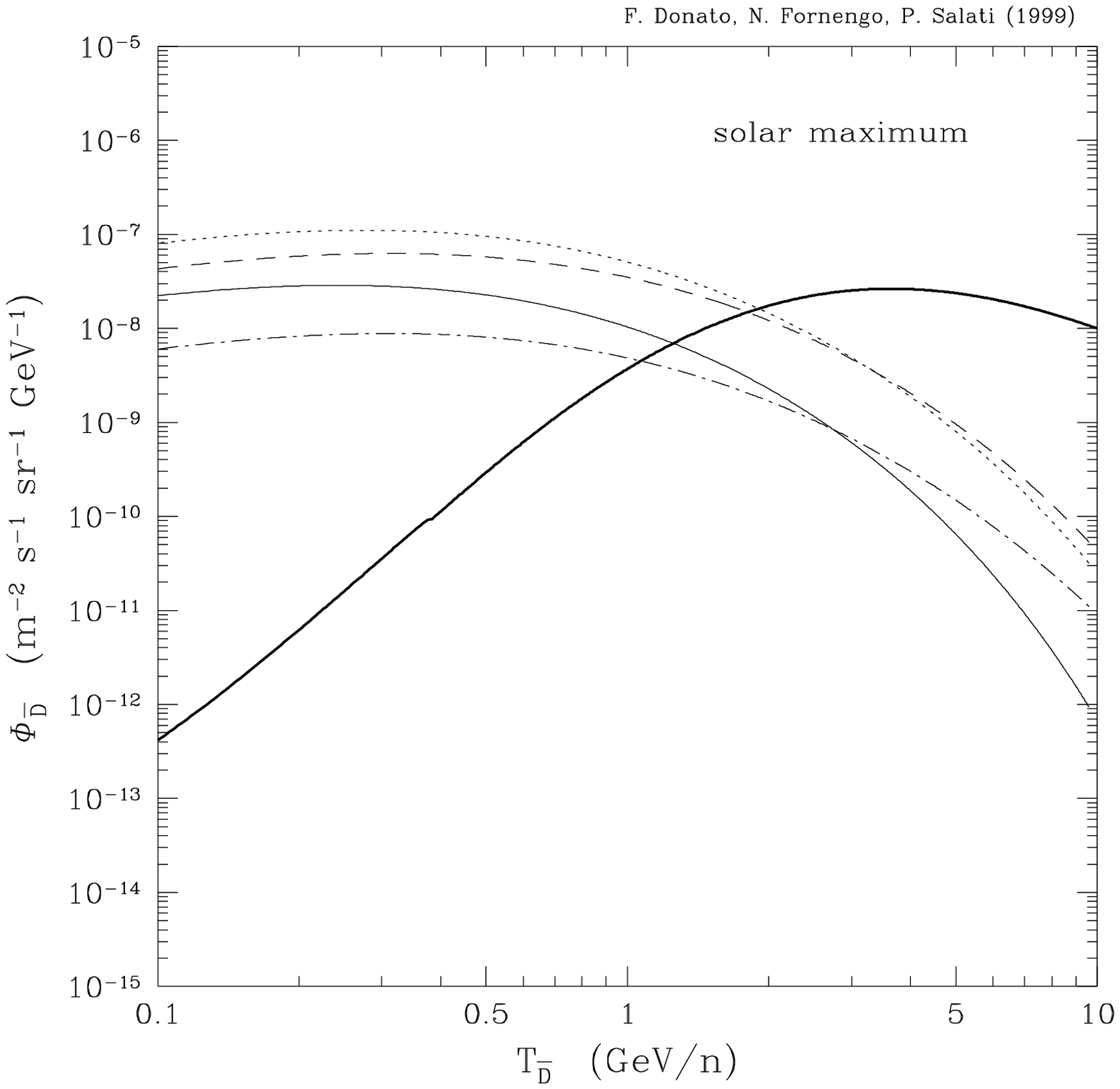}}}
{\resizebox{0.5\textwidth}{!}
{\includegraphics*[1.5cm,6.5cm][18.5cm,23.cm]{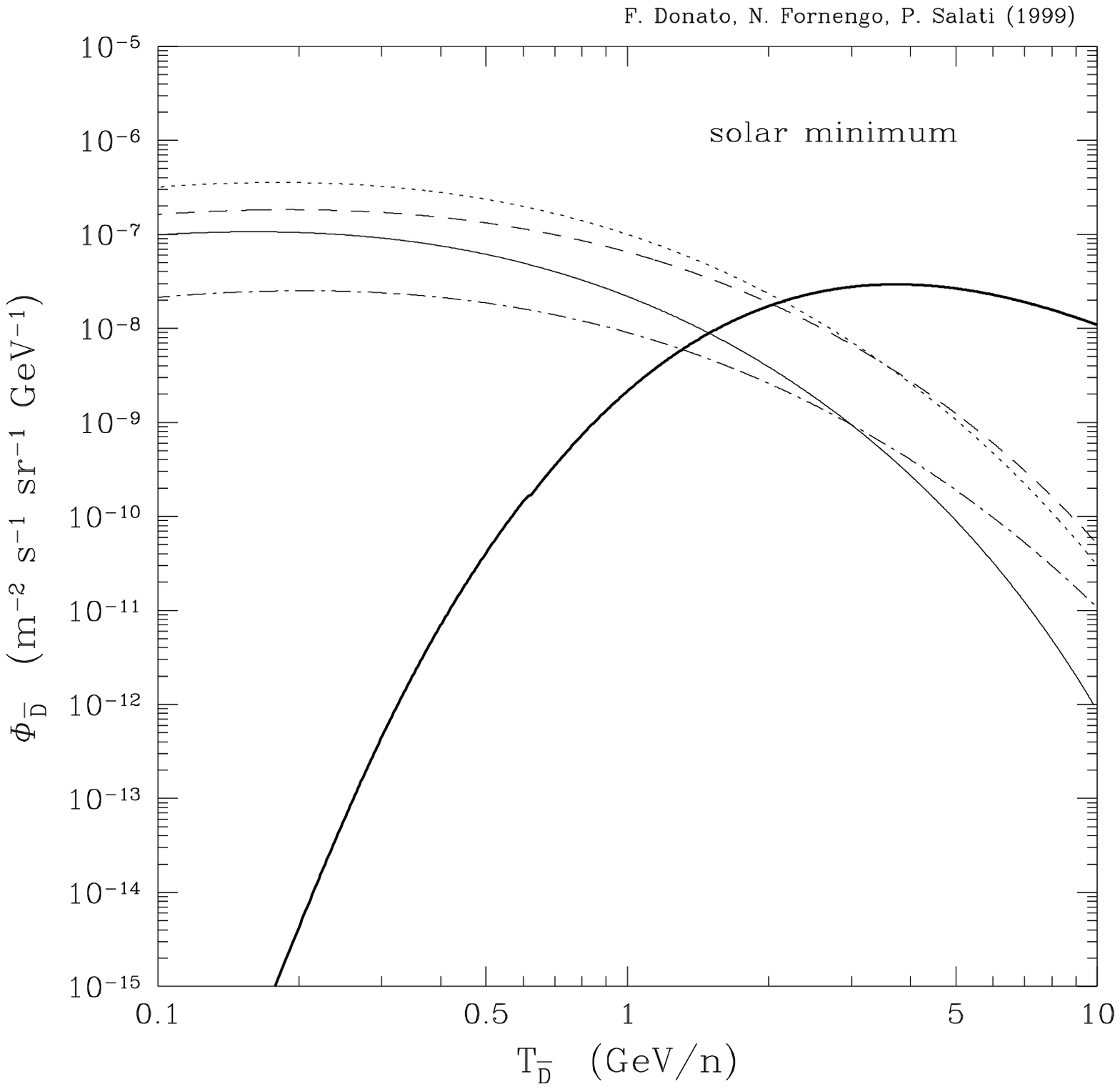}}}
}
\caption{
Same as in Fig.~\ref{fig:dbar_prim_is} but modulated at solar
maximum ({\rm left}) and minimum ({\rm right}).
}
\label{fig:dbar_prim_solmod}
\end{figure*}

\begin{figure*}[h!]
\centerline{
\resizebox{0.7\textwidth}{!}
{\includegraphics*[1.5cm,6.5cm][18.5cm,23.cm]{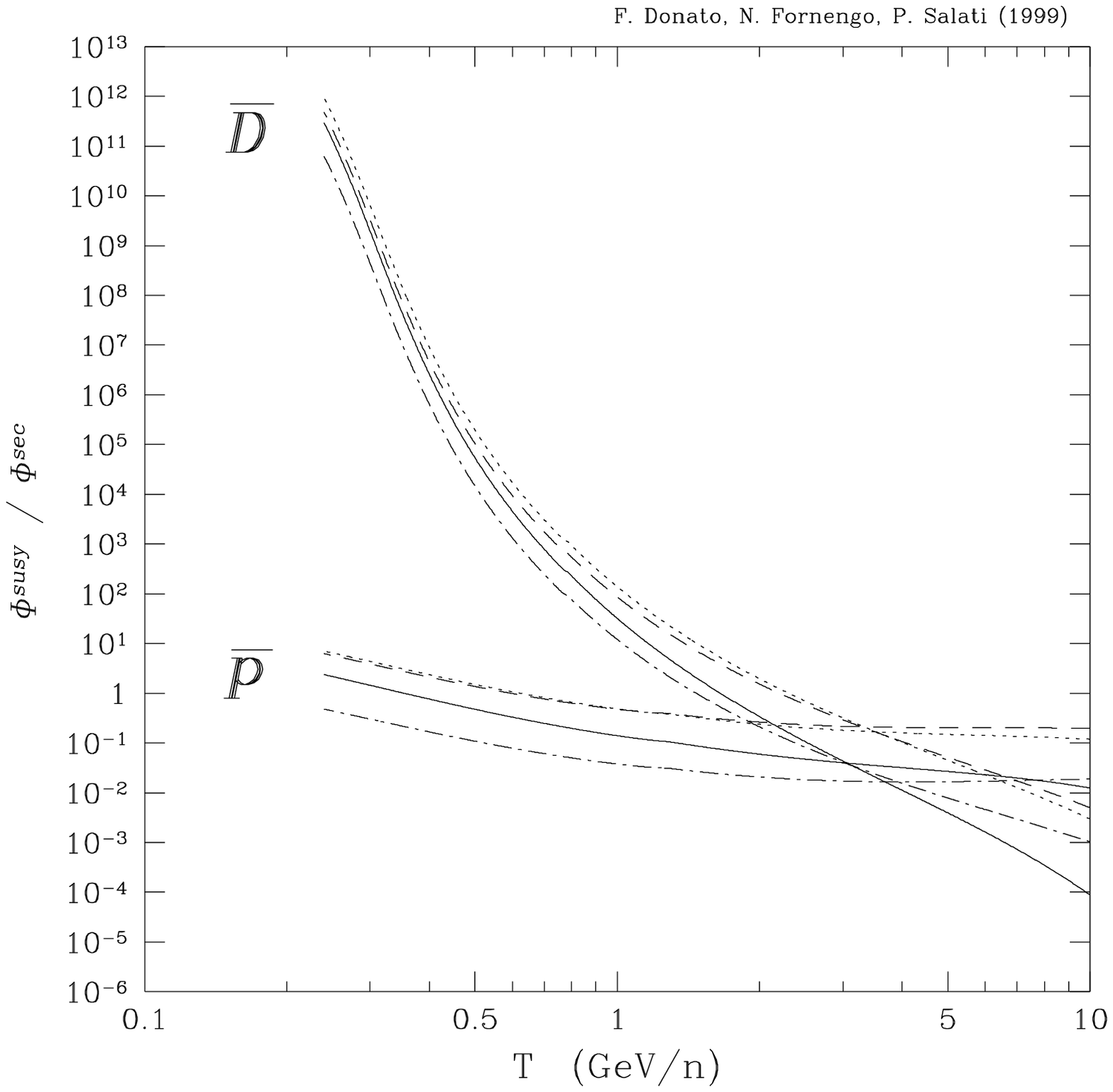}}
}
\caption{
The supersymmetric--to--secondary IS flux ratio for antiprotons
(lower curves) and antideuterons (upper curves) is presented
as a function of the kinetic energy per nucleon. The supersymmetric
configurations are those reported in table~\ref{table:susy} and
featured in Figs.~\ref{fig:dbar_prim_is} and
\ref{fig:dbar_prim_solmod}. Below a few GeV/n, the flux ratio is
always larger for {\dbar}'s than for {\pbar}'s.
For the supersymmetric configurations of table~\ref{table:susy}, the
antiproton signal is swamped into its background. This is not the case for
antideuterons. At low energy, the flux of primaries is several orders of
magnitude above the {\dbar} background.
}
\label{fig:susy_over_sec}
\end{figure*}

As a theoretical framework, we use the Minimal Supersymmetric extension of the
Standard Model (MSSM) \cite{susy}, which conveniently  describes the
supersymmetric phenomenology at the electroweak scale, without too strong
theoretical assumptions. This model has been largely adopted by many
authors for evaluations of the neutralino relic abundance and detection rates
(for reviews, see Refs.~\cite{ICTP,jkg}).

The MSSM is defined at the electroweak scale as a straightforward supersymmetric
extension of the Standard Model. The Higgs sector consists of two Higgs doublets
$H_1$ and $H_2$ and, at the tree level, is fully described by two free parameters,
namely: the ratio of the two vacuum expectation values
$\tan\beta  \equiv \langle H_2 \rangle/\langle H_1\rangle$ and the mass of one
of the three neutral physical Higgs fields, which we choose to be the mass $m_A$ of
the neutral pseudoscalar one. Once radiative corrections are introduced, the Higgs
sector depends also on the squark masses through loop diagrams. The radiative
corrections to the neutral and charged Higgs bosons, adopted in the present paper,
are taken from Refs.~\cite{carena,haber}.
The other parameters of the model are defined in the superpotential, which contains
all the Yukawa interactions and the Higgs--mixing term $\mu H_1 H_2$, and  in the
soft--breaking Lagrangian, which contains the trilinear and bilinear  breaking parameters
and the soft gaugino and scalar mass  terms.
In this model, the neutralino is defined as the lowest--mass linear superposition of
photino ($\tilde \gamma$), zino ($\tilde Z$) and the two higgsino states
($\tilde H_1^{\circ}$, $\tilde H_2^{\circ}$)
\beq
\chi \equiv a_1 \tilde \gamma + a_2 \tilde Z + a_3 \tilde H_1^{\circ}
+ a_4 \tilde H_2^{\circ} \;\; .
\label{eq:neu}
\eeq
In order to deal with manageable models, it is necessary to introduce some
assumptions which establish relations among the too many free parameters
at the electroweak scale. We adopt the following usual conditions.
All trilinear parameters are set to zero except those of the third family,
which are unified to a common value $A$.
All squarks and sleptons soft--mass parameters are taken as
degenerate: $m_{\tilde l_i} = m_{\tilde q_i} \equiv m_0$.
The gaugino masses are assumed to unify at $M_{GUT}$, and this implies that
the $U(1)$ and $SU(2)$ gaugino masses are related at the electroweak scale
by $M_1= (5/3) \tan^2 \theta_W M_2$.
When all these conditions are imposed, the supersymmetric parameter space is
completely described by six independent parameters, which we choose to be:
$M_2, \mu, \tan\beta, m_A, m_0, A$. In our analyses, we vary them in the
following ranges:
$20\;\mbox{GeV} \leq M_2 \leq  500\;\mbox{GeV}$;
$20\;\mbox{GeV} \leq |\mu| \leq  500\;\mbox{GeV}$;
$80\;\mbox{GeV} \leq m_A \leq  1000\;\mbox{GeV}$;
$100\;\mbox{GeV} \leq m_0 \leq  1000\;\mbox{GeV}$;
$-3 \leq {\rm A} \leq +3$;
$1 \leq \tan \beta \leq 50$.

The supersymmetric parameter space is constrained by all the experimental limits
achieved at accelerators on supersymmetric and Higgs searches \cite{lep189}. Also
the constraints due to the $b \rightarrow s + \gamma$ process \cite{glenn,barate}
have been taken into account (see Ref.~\cite{noi3} for a discussion of 
our implementation of the $b \rightarrow s + \gamma$ constraint and 
for the relevant references).
We further require the neutralino to be the Lightest Supersymmetric Particle
(LSP)  and the  supersymmetric configurations to provide a neutralino relic
abundance in accordance with the cosmological bound
$\Omega_{\chi}h^2 \leq 0.7$ \cite{ICTP}.

For the evaluation of the averaged annihilation cross section
$<\sigma_{\rm ann} v>$, we have followed the procedure outlined in
Ref.~\cite{bffm}. We have considered all the tree--level diagrams which are
responsible for neutralino annihilation and which are relevant to $\pbar$
production, namely: annihilation into quark--antiquark pairs, into gauge bosons,
into a Higgs boson pair and into a Higgs and a gauge boson. For each final state
we have considered all the relevant Feynman diagrams, which involve the
exchange of Higgs and $Z$ bosons in the s--channel and the exchange of squarks,
neutralinos and charginos in the t and u--channels. Finally, we have included
the one--loop diagrams which produce a two--gluon final state \cite{bu}.
The $\pbar$ differential distribution
${dN_{\pbar}} / {d \Epbar}$ has been evaluated as discussed in Ref.~\cite{bffm}.
Here we only recall that we have calculated the branching ratios
$B_{\rm \chi h}^{\rm (F)}$ for all annihilation final states F which may
produce $\pbar$'s. These final states fall into two categories~:
(i) direct production of quarks and gluons and
(ii) generation of quarks through the intermediate production of Higgs bosons,
gauge bosons and $t$ quark.
In order to obtain the distributions ${dN_{\pbar}^{\rm h}} / {d \Epbar}$,
the hadronization of quarks and gluons has been computed by using the Monte
Carlo code Jetset 7.2 \cite{jet}. For the top quark, we have considered it to decay
before hadronization.
The source term for supersymmetric antideuterons
\beq
q_{\dbar}^{\rm susy} \left( \chi + \chi \to \dbar + \ldots \right)
\; = \; 
<\sigma_{\rm ann} v> \,
{\displaystyle \frac {dN_{\dbar}}{d \Edbar}} \,
\left\{ {\displaystyle \frac{\rho_{\chi}}{\mC}} \right\}^{2}
\eeq
supplements the spallation contribution $q_{\dbar}^{\rm sec}$ in
the diffusion Eq.~(\ref{diffusion_dbar_disk}). The propagation of
primary antideuterons from the remote regions of the galactic halo
to the Earth has been treated as explained in Ref.~\cite{bottino98}.
The neutralino distribution has been assumed to be spherical, with
radial dependence
\beq
\rho_{\chi} \; = \;
\rho_{\chi}^{\odot} \,
\left\{
{\displaystyle \frac{a^{2} + r_{\odot}^{2}}{a^{2} + m^{2}}}
\right\} \;\; ,
\eeq
where $m^{2} = r^{2} + z^{2}$. The solar system is at a distance
$r_{\odot}$ of 8 kpc from the galactic center. The dark matter
halo has a core radius $a = 3.5$ kpc and its density in the solar
neighborhood is $\rho_{\chi}^{\odot} = 0.4$ GeV cm$^{-3}$
\cite{ICTP}.

In Fig.~\ref{fig:dbar_prim_is}, both primary (supersymmetric) and
secondary (spallation) interstellar antideuterons energy spectra are
presented. The secondary flux (heavier solid line) drops sharply at
low energies as discussed above. The four supersymmetric examples of
table~\ref{table:susy} are respectively featured by the solid
(a), dotted (b), dashed (c) and dot-dashed (d) curves. The corresponding
primary fluxes flatten at low energy where they reach a maximum.
As the secondary {\dbar} background vanishes, the supersymmetric
signal is the largest. Neutralino annihilations actually take place at rest
in the galactic frame. The fragmentation and subsequent hadronization of
the jets at stake tend to favour the production of low--energy species.
Therefore, the spectrum of supersymmetric antiprotons -- and antineutrons
-- is fairly flat below $\sim$ 1 GeV. For the same reasons, the coalescence of
the primary antideuterons produced in neutralino annihilations
predominantly takes place with the two antinucleons at rest, hence a flat
spectrum at low energy, as is clear in Fig.~\ref{fig:dbar_prim_is}. The fusion
of an antideuteron requires actually that its antinucleon constituents should
be aligned in momentum space. Consequently, secondary antideuterons are
completely depleted below $\sim 1$ GeV while the primary species are mostly
produced in that low--energy regime.
This trend still appears once the energies and fluxes are modulated.
The left and right panels of Fig.~\ref{fig:dbar_prim_solmod} respectively
show the effects of solar modulation at maximum and minimum. The 
spallation background somewhat flattens. It is still orders of magnitude below
the supersymmetric signal which clearly exhibits a plateau.

It is difficult to establish a correlation between the {\dbar} flux
and the neutralino mass. In case (c), for instance, $\mC$ is $\sim$ 3 times
larger as in case (a) and yet the corresponding antideuteron flux is larger.
It is not obvious either that gaugino--like mixtures lead to the largest
{\dbar} signals. Table~\ref{table:susy} gives a flavor of the complexity
and of the richness of the supersymmetric parameter space.

In Fig.~\ref{fig:susy_over_sec}, the supersymmetric--to--spallation IS flux ratio
for antiprotons (lower curves) and antideuterons (upper curves) are presented
as a function of the kinetic energy per nucleon. In the case of antiprotons, the 
primary--to--secondary ratio is much smaller than for antideuterons.
For the configurations of table~\ref{table:susy} presented here, the {\pbar} primary
flux is at the same level as the spallation background. The supersymmetric
antiproton signal is swamped in the flux of the secondaries. This is not the case for
antideuterons. At low energies, their supersymmetric flux is several orders of
magnitude above background. Antideuterons appear therefore as a much cleaner
probe of the presence of supersymmetric relics in the galactic halo than antiprotons.
The price to pay however is a much smaller flux.
Typical {\dbar} spectra may reach up to $10^{-6}-10^{-5} \, \DFLUX$. This 
corresponds to an antiproton signal of $10^{-2}-10^{-1} \, \DFLUX$, \ie, four orders
of magnitude larger. It is therefore crucial to ascertain which portion of the
supersymmetric configurations will be accessible to future experiments
through the detection of low--energy cosmic--ray antideuterons.

\section{Discussion and conclusions.}
\label{sec:conclusion}

In order to be specific, we have estimated the amount of antideuterons
which may be collected by the AMS experiment once it is on board ISSA.
The future space station is scheduled to orbit at 400 km above sea level,
with an inclination of $\alpha = 52^{\circ}$ with respect to the Earth
equator. A revolution takes about 1.5 hours so that ISSA should fly over
the same spot every day. The AMS detector may be pictured as a 
cylindrical magnetic field with diameter $D = 110$ cm. At any time, its axis 
points towards the local vertical direction. The colatitude of the north
magnetic pole has been set equal to $\Upsilon = 11^{\circ}$. At any given
time $t$ along the orbit, the geomagnetic latitude $\varrho$ of ISSA may be
inferred from
\bea
\sin \varrho & = &
\sin \Upsilon \,
\cos \left( \Omega_{\rm sid} t \right) \,
\cos \left( \Omega_{\rm orb} t + \varphi \right) +
\\
& + &
\cos \alpha \, \sin \Upsilon \,
\sin \left( \Omega_{\rm sid} t \right) \,
\sin \left( \Omega_{\rm orb} t + \varphi \right) \; + \;
\sin \alpha \, \cos \Upsilon \,
\sin \left( \Omega_{\rm orb} t + \varphi \right) \;\; ,
\nonumber
\ena
where $\Omega_{\rm sid}$ and $\Omega_{\rm orb}$ respectively
denote the angular velocities associated to the sideral rotation of the
Earth and to the orbital motion of the space station. The phase
$\varphi$ depends on the orbital initial conditions and does not affect
the result if a large number of revolutions -- typically 100 -- is considered.
The Earth is shielded from cosmic--rays because its magnetic field
prevents particles from penetrating downwards. At any
given geomagnetic latitude $\varrho$, there exists a rigidity cut-off
${\cal R}_{\rm min}$ below which the cosmic--ray flux is suppressed.
This lower bound depends on the radius $R$ of the orbit through
\beq
{\cal R}_{\rm min} \; = \;
{\displaystyle \frac{\mu_{\oplus}}{R^{2}}} \,
{\displaystyle \frac{\cos^{4} \varrho}{\varpi^{2}}} \;\; ,
\label{geomagnetic_1}
\eeq
where $\mu_{\oplus}$ denotes the Earth magnetic dipole moment
and ${\mu_{\oplus}} / {R_{\oplus}^{2}} \simeq 60$ GV. The term
$\varpi$ stands for
\beq
\varpi \; = \; 1 \, + \, \sqrt{1 + \cos \theta \, \cos^{3} \varrho} \;\; .
\label{geomagnetic_2}
\eeq
It depends on the angle $\theta$ between the cosmic--ray
momentum at the detector and the local east--west line that points
in the ortho--radial direction of an axisymmetric coordinate system.
Notice that because we are interested here in singly charged species, the
rigidity amounts to the momentum $p$.
Once the cosmic--ray energy as well as the geomagnetic latitude are
specified, the solid angle $\Omega_{\rm cut}$ inside which the
direction of the incoming particle lies may be derived from
relations~(\ref{geomagnetic_1}) and (\ref{geomagnetic_2}).
The AMS detector looks upwards within $\sim 27^{\circ}$ around
the vertical. This corresponds to a solid angle of
$\Omega_{\rm det} = 0.68$ sr. Because the apparatus does not point
towards the local east or west, impinging particles may not be seen by
the  instrument. The effective solid angle
$\Omega_{\rm eff}$ through which they are potentially detectable
corresponds to the overlap, if any, between $\Omega_{\rm cut}$ and
$\Omega_{\rm det}$. The value of $\Omega_{\rm eff}$ depends on
the cosmic--ray rigidity $p$ as well as on the precise location of the detector
along the orbit.
The detector acceptance may therefore be defined as
\beq
\aleph \left( p \right) \; = \;
{\displaystyle \frac{\pi}{4}} D^{2} \,
{\displaystyle \int} \Omega_{\rm eff} \left( p , t \right) \, dt \;\;,
\eeq
where the time integral runs over the duration $\tau$ of the space
mission. In the case of AMS on board ISSA, $\tau$ is estimated to be
$10^{8}$ s (3 yrs).
Between 100 MeV/n and 100 GeV/n, we infer a total acceptance of
$5.8 \times 10^{9}$ ${\rm m^{2} \; s \; sr \; GeV}$
for antiprotons and of
$6 \times 10^{9}$ ${\rm m^{2} \; s \; sr \; GeV}$ for
antideuterons.
The net number of cosmic--ray species which AMS may collect on
board ISSA is actually a convolution of the detector acceptance with
the relevant differential flux at Earth. For antideuterons, this leads to
\beq
N_{\dbar} \; = \;
{\displaystyle \int} \aleph \left( p_{\dbar}^{\oplus} \right) \,
\Phi_{\dbar}^{\oplus} \, dT_{\dbar}^{\oplus} \;\; ,
\label{dbar_convolution}
\eeq
where the integral runs on the {\dbar} modulated energy $T_{\dbar}^{\oplus}$.

Integrating the secondary flux discussed in Sect.~\ref{sec:secondary}
leads respectively to a total of 12.3 and 13.4 antideuterons, depending on
whether the solar cycle is at maximum or minimum. These spallation
{\dbar}'s are mostly expected at high energies. As is clear from
Figs.~\ref{fig:dbar_prim_is} and \ref{fig:dbar_prim_solmod}, the secondary
flux drops below the supersymmetric signal below a few GeV/n. The transition
typically takes place for an interstellar energy of 3 GeV/n. Below that value,
the secondary antideuteron signal amounts to a total of only 0.6 (solar
maximum) and 0.8 (solar minimum) nuclei.
Most of the supersymmetric signal is therefore concentrated in a low--energy
band extending from the AMS threshold of 100 MeV/n up to a modulated
energy of 2.6 GeV/n (maximum) or 2.84 GeV/n (minimum) which corresponds
to an upper bound of 3 GeV/n in interstellar space. In this low--energy
region where spallation antideuterons yield a negligible background, 
the AMS acceptance is
$2.2 \times 10^{7}$ ${\rm m^{2} \; s \; sr \; GeV}$ for antiprotons and
$5.5 \times 10^{7}$ ${\rm m^{2} \; s \; sr \; GeV}$ for antideuterons.

\begin{figure*}[h!]
\centerline{
\resizebox{0.7\textwidth}{!}
{\includegraphics*[1.5cm,6.5cm][18.5cm,23.cm]{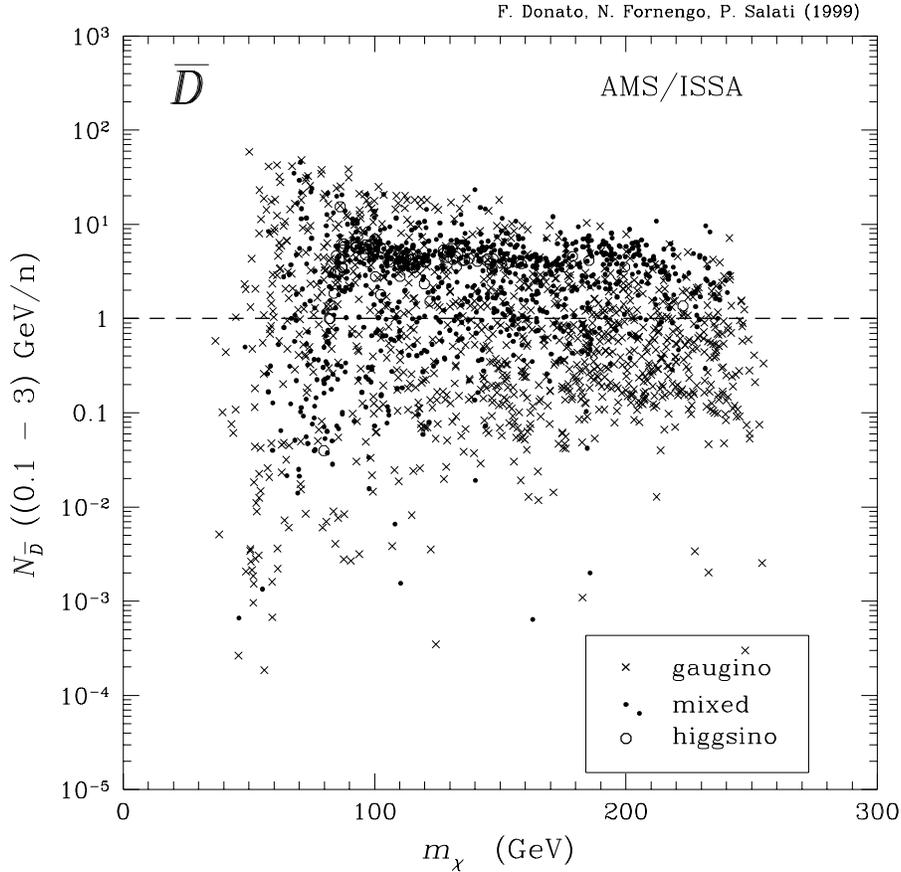}}
}
\caption{
The supersymmetric {\dbar} flux has been integrated over
the range of IS energies extending from 0.1 up to 3 GeV/n.
The resulting yield $N_{\dbar}$ of antideuterons which AMS
on board ISSA can collect is plotted as a function of the neutralino
mass $\mC$. Modulation has been considered at solar maximum.
}
\label{fig:ndbar_versus_mchi}
\end{figure*}

For each supersymmetric configuration, the {\dbar} flux has
been integrated over that low--energy range. The resulting yield $N_{\dbar}$
which AMS may collect on board ISSA is presented as a function of the
neutralino mass $\mC$ in the scatter plot of Fig.~\ref{fig:ndbar_versus_mchi}.
During the AMS mission, the solar cycle will be at maximum. Most of 
the configurations are gaugino like (crosses) or mixed combinations of
gaugino and higgsino states (dots). A significant portion of the parameter
space is associated to a signal exceeding one antideuteron -- horizontal
dashed line. In a few cases, AMS may even collect more than a dozen of
low--energy {\dbar} nuclei. However, when the antideuteron signal exceeds
$\sim$ 20 particles, the associated antiproton flux is larger than what
BESS 95 + 97 \cite{bess_2} has measured.

\begin{figure*}[h!]
\centerline{
{\resizebox{0.5\textwidth}{!}
{\includegraphics*[1.5cm,6.5cm][18.5cm,23.cm]{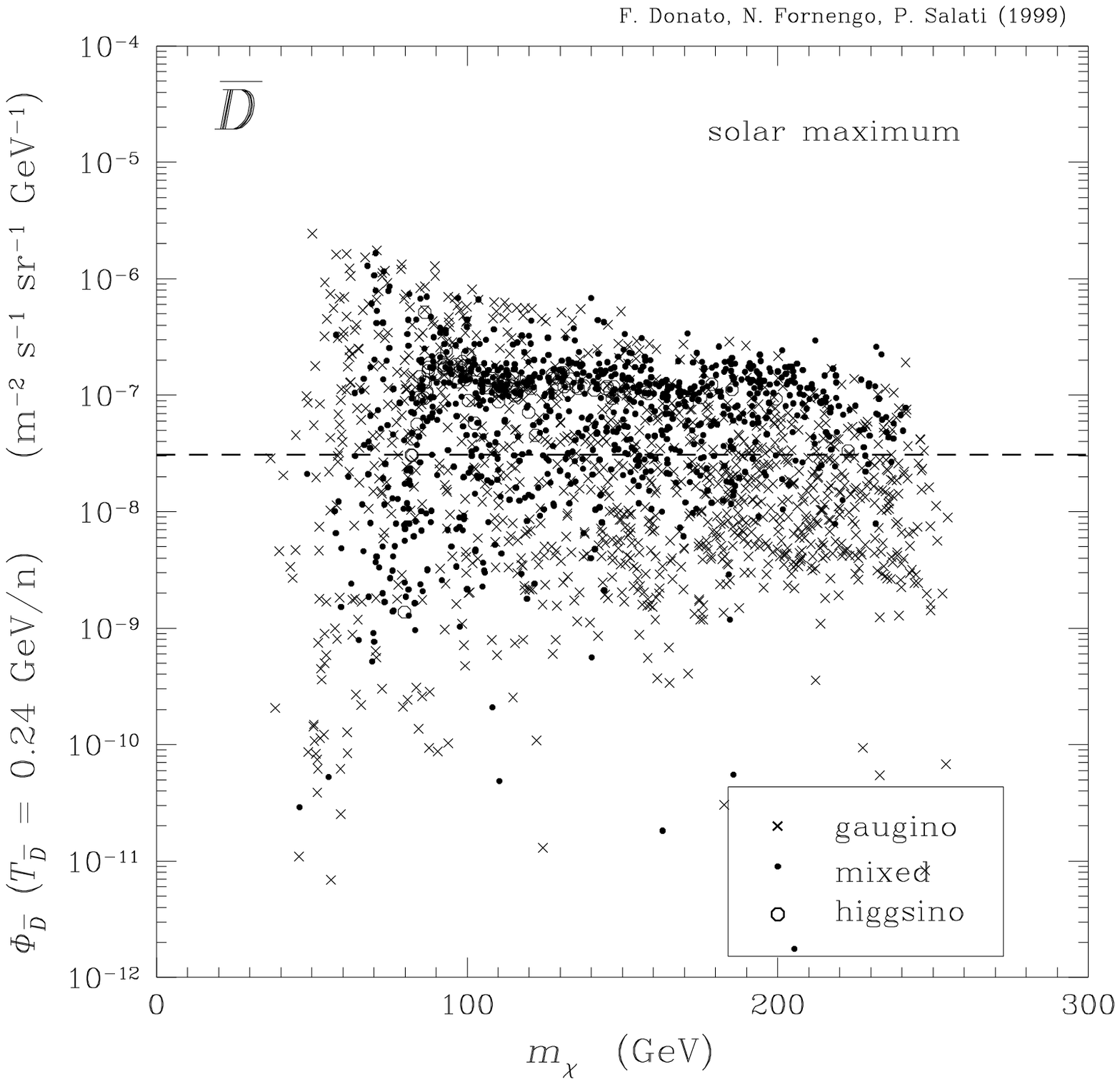}}}
{\resizebox{0.5\textwidth}{!}
{\includegraphics*[1.5cm,6.5cm][18.5cm,23.cm]{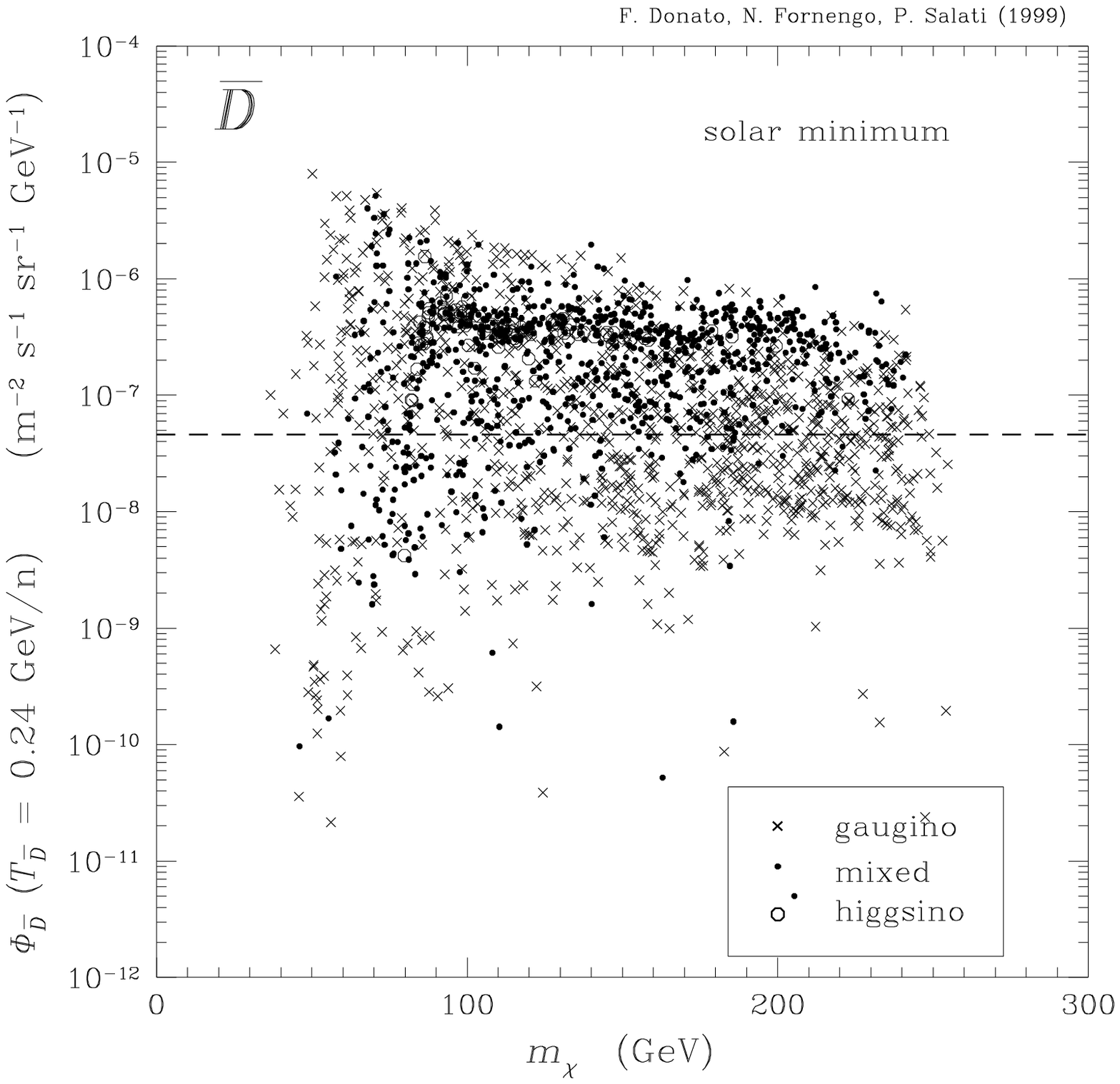}}}
}
\caption{
Scatter plots in the plane $m_{\chi}$--$\Phi_{\dbar}^{\oplus}$. The Earth
antideuteron flux $\Phi_{\dbar}^{\oplus}$ has been computed at solar
maximum ({\rm left}) and minimum ({\rm right}), for a modulated
energy of 0.24 GeV/n. Configurations lying above the horizontal lines
correspond to the detection of at least one antideuteron in the range
of interstellar energies 0.1 -- 3 GeV, by an experiment of the AMS caliber
on board ISSA.
}
\label{fig:dbar_scatter_solmod}
\end{figure*}

The scatter plot of Fig.~\ref{fig:ndbar_versus_mchi} may be translated
into a limit on the antideuteron flux $\Phi_{\dbar}^{\oplus}$ at the
Earth. Table~\ref{table:susy} gives a flavor of the relation between that
flux and the yield $N_{\dbar}$ of low--energy antideuterons. At solar
maximum, a value of $N_{\dbar} = 1$ translates, on average, into a flux
of $\sim 3.2 \times 10^{-8}$ {\dbar} {\DFLUX} for a modulated energy
of 240 MeV/n. The energy spectrum matters of course. For the steep
differential flux of case (a), a value of
$4.8 \times 10^{-8}$ {\dbar} {\DFLUX} is necessary in order to achieve a
signal of at least one antideuteron. In case (d) where the spectrum is much
flatter, the same {\dbar} yield is reached for a flux of only
$2.8 \times 10^{-8}$ {\dbar} {\DFLUX}. The horizontal dashed lines
of Figs.~\ref{fig:dbar_scatter_solmod} should therefore be understood
as averaged limits. They are nevertheless indicative of the level of
sensitivity which may be reached through the search for low--energy
antideuterons. The left and right panels respectively correspond to a
solar activity taken at maximum and minimum. In these scatter plots,
the {\dbar} modulated flux is featured as a function of the neutralino
mass $\mC$. The antideuteron energy at the Earth has been set equal
to 240 MeV/n. The flux $\Phi_{\dbar}^{\oplus}$ is larger at solar
minimum -- when modulation is weaker --than at maximum. The
lower the cosmic--ray energy, the larger that effect. The plateaux of
Figs.~\ref{fig:dbar_prim_solmod} illustrate the flatness of the 
supersymmetric {\dbar} spectra at low energies. These plateaux actually
exhibit a shift by a factor $\sim$ 3 between the left and right panels.
Accordingly, the constellation of supersymmetric configurations
in Figs.~\ref{fig:dbar_scatter_solmod} is shifted upwards, by the same
amount, between solar maximum (left panel) and minimum (right
panel).
At larger energies, the variation of the flux at Earth during
the solar cycle is milder. Above a few GeV/n,
solar modulation has no effect. The number of supersymmetric
antideuterons collected at low energy obtains from the convolution
of Eq.~(\ref{dbar_convolution}). It also varies during the solar cycle,
in a somewhat lesser extent however than the above mentioned plateaux.
Between maximum and minimum, the value of $N_{\dbar}$ only
varies by a factor of $\sim$ 2, to be compared to a flux increase of
$\sim$ 3. At solar maximum, when AMS/ISSA will be operating,
a signal of one antideuteron translates into a flux sensitivity of
$\sim 3.2 \times 10^{-8}$ antinuclei {\DFLUX}. At minimum, the
same signal would translate into the weaker limit of
$\sim 4.8 \times 10^{-8}$ antideuterons {\DFLUX} and the horizontal
dashed line is shifted upwards by $\sim$ 50\%. The supersymmetric
configurations which an antideuteron search may unravel are
nevertheless more numerous at solar minimum. Between the left and
the right panels, the constellation of representative points is actually
shifted upwards and, relative to the limit of sensitivity, the increase
amounts to a factor $\sim$ 2.
In spite of the low fluxes at stake, the antideuteron channel is sensitive
to a respectable number of supersymmetric configurations.

\begin{figure*}[h!]
\centerline{
\resizebox{0.7\textwidth}{!}
{\includegraphics*[1.5cm,6.5cm][18.5cm,23.cm]{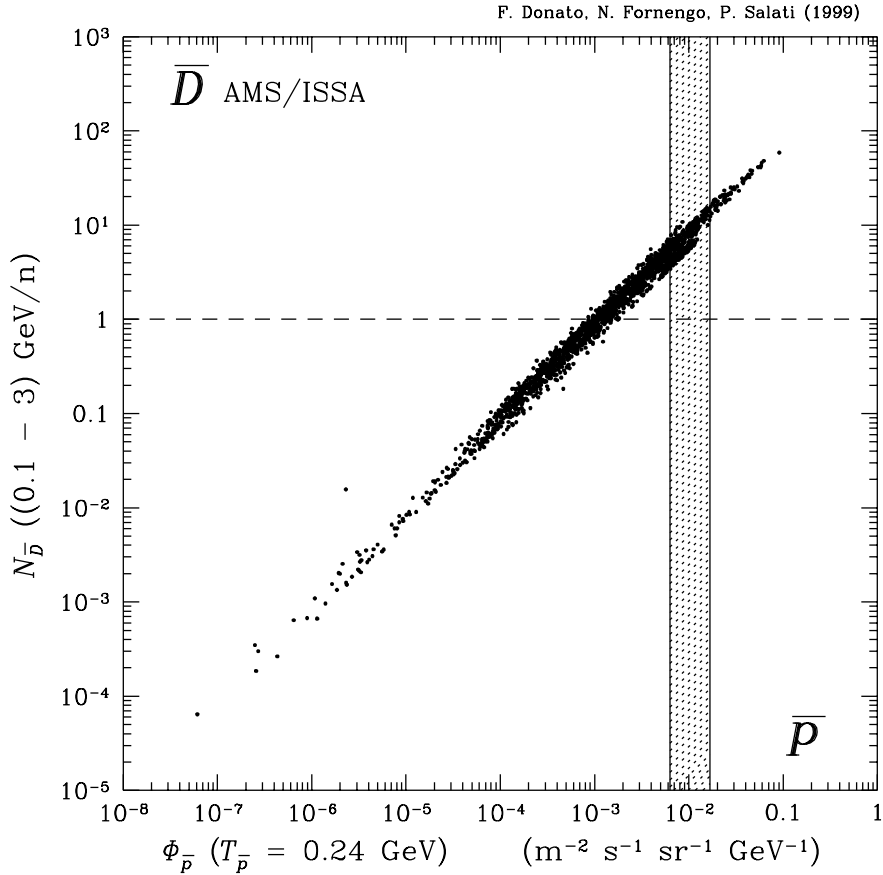}}
}
\caption{
In this scatter plot, the antideuteron yield $N_{\dbar}$ of
Fig.~\ref{fig:ndbar_versus_mchi} is featured against the supersymmetric
{\pbar} flux. The antideuteron signal is estimated at solar maximum.
This corresponds to the AMS mission on board the space station. The
{\pbar} flux is derived on the contrary at solar minimum, in the same
conditions as the BESS 95 + 97 flights [26] whose combined
measurements are indicated by the vertical shaded band for a {\pbar}
energy of 0.24 GeV. The correlation between the antiproton and
antideuteron signals is strong.
}
\label{fig:ndbar_pbar_scatter}
\end{figure*}

\begin{figure*}[h!]
\centerline{
\resizebox{0.7\textwidth}{!}
{\includegraphics*[1.5cm,6.5cm][18.5cm,23.cm]{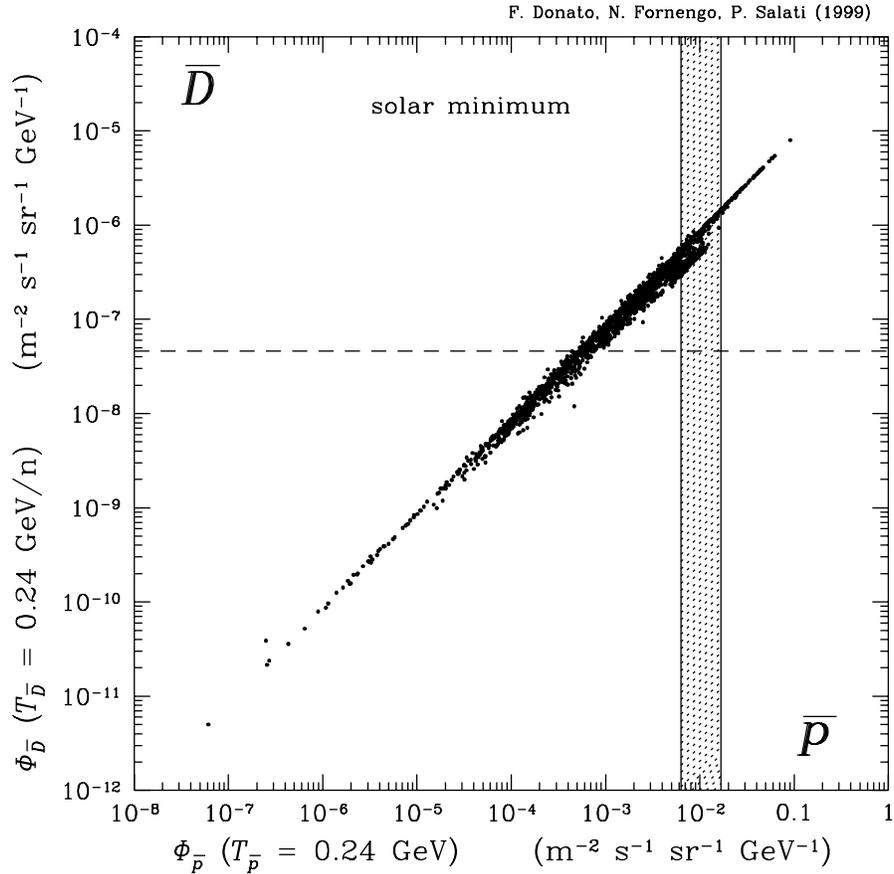}}
}
\caption{
Both supersymmetric antideuteron and antiproton fluxes at
the Earth are plotted against each other. They are modulated
at solar minimum, while the energy per nucleon is
$T_{\dbar}^{\oplus} / {\rm n} = 0.24$ GeV/n. As in
Fig.~\ref{fig:ndbar_pbar_scatter}, the configurations are clearly
aligned, hence a strong correlation between the antiproton and
antideuteron signals.
}
\label{fig:dbar_pbar_scatter}
\end{figure*}

Supersymmetric antiprotons are four orders of magnitude more abundant
in cosmic--rays than antideuterons -- see table~\ref{table:susy}.
However, as already discussed, they may be swamped in the background
arising from the secondaries. The AMS experiment will collect a large 
number of antiprotons on board ISSA. Our concern is whether an 
hypothetical supersymmetric {\pbar} signal may be disentangled from
the background. Because the latter still suffers from large theoretical
uncertainties, we are afraid that antiproton searches in cosmic--rays are
not yet the ultimate probe for the existence of supersymmetric relics in
the Milky Way. As discussed in Refs.~\cite{bottino98,bergstrom99,bieber99},
the distribution of secondary antiprotons turns out to be flatter than
previously estimated. Therefore, it is still a quite difficult task to
ascertain which fraction of the measured antiproton spectrum may
be interpreted as a supersymmetric component. Notice however
that as soon as the secondary {\pbar} flux is reliably estimated,
low--energy antiproton searches will become a more efficient tool.
Meanwhile, we must content ourselves with using observations
as a mere indication of what a supersymmetric component cannot
exceed. The vertical shaded band of
Figs.~\ref{fig:ndbar_pbar_scatter} and \ref{fig:dbar_pbar_scatter}
corresponds actually to the 1--$\sigma$ antiproton flux which the
BESS 95 + 97 experiments \cite{bess_2} have measured at a {\pbar}
kinetic energy of 0.24 GeV.
In Fig.~\ref{fig:ndbar_pbar_scatter}, the supersymmetric
antideuteron yield $N_{\dbar}$ has been derived at solar
maximum. This corresponds to the conditions of the future
AMS mission on board the space station. The antideuteron
yield is plotted as a function of the associated supersymmetric
{\pbar} flux at Earth. The latter is estimated at solar minimum
to conform to the BESS data to which the vertical band refers.
The scatter plot of Fig.~\ref{fig:ndbar_pbar_scatter} illustrates the
strong correlation between the antideuteron and the antiproton
signals, as may be directly guessed from
Eq.~\ref{dNdbar_on_dEdbar_susy}. The horizontal dashed line
indicates the level of sensitivity which AMS/ISSA may reach.
Points located above that line but on the left of the shaded vertical
band are supersymmetric configurations that are not yet excluded
by antiproton searches and for which the antideuteron yield is
potentially detectable. The existence of such configurations illustrates
the relevance of an antideuteron search at low energies.
As shown in Fig.~\ref{fig:dbar_pbar_scatter}, the number
of interesting configurations is largest at solar minimum.
Both {\dbar} and {\pbar} fluxes at Earth are plotted against
each other. Energies have been set equal to 0.24 GeV/n.
The correlation between the antideuteron and antiproton
cosmic--ray fluxes is once again noticeable.

Once the energy spectrum of the secondary component
is no longer spoilt by considerable theoretical uncertainties,
measurements of the antiproton cosmic--ray flux will
be a powerful way to test the existence of supersymmetric
relics in the galactic halo. In the mean time, searches for
low--energy antideuterons appear as a plausible alternative, worth
being explored. A dozen spallation antideuterons should be
detected by the future AMS experiment on board ISSA
above a few GeV/n. For energies less than $\sim$ 3 GeV/n,
the {\dbar} spallation component becomes negligible and may
be supplanted by a potential supersymmetric signal. We
conclude that the discovery of a few low--energy antideuterons
should be taken seriously as a clue for the existence of
massive neutralinos in the Milky Way.

\vskip 1.cm
\noindent {\bf Acknowledgements}
\vskip 0.5cm
\noindent We would like to express our gratitude toward S.~Bottino for
stimulating discussions. We also wish to thank R.~Battiston and
J.P.~Vialle for supplying us with useful information pertaining to
the AMS experiment. This work was supported by DGICYT under grant
number PB95--1077 and by the TMR network grant ERBFMRXCT960090 of the
European Union.

\end{document}